# Accurate Virus Identification with Interpretable Raman Signatures by Machine Learning

Jiarong Ye[1], Yin-Ting Yeh[2], Yuan Xue[3], Ziyang Wang[4], Na Zhang[2], He Liu[2], Kunyan Zhang[4], RyeAnne Ricker[5], Zhuohang Yu[2], Allison Roder[6], Nestor Perea Lopez[2], Lindsey Organtini[7], Wallace Greene[8], Susan Hafenstein[7], Huaguang Lu[9], Elodie Ghedin[6], Mauricio Terrones[2], Shengxi Huang[4], Sharon Xiaolei Huang[1,*]

[1.] College of Information Sciences and Technology, The Pennsylvania State University, University Park, PA, USA
[2.] Department of Physics, The Pennsylvania State University, University Park, PA, USA
[3.] Department of Electrical and Computer Engineering, Johns Hopkins University, Baltimore, MD, USA
[4.] Department of Electrical and Computer Engineering, The Pennsylvania State University, University Park, PA, USA
[5.] Department of Biomedical Engineering, George Washington University, Washington, D.C., USA
[6.] Systems Genomics Section, Laboratory of Parasitic Diseases, National Institute of Allergy and Infectious Diseases, National Institutes of Health, Bethesda, MD, USA
[7.] Department of Biochemistry and Molecular Biology, The Pennsylvania State University, University Park, PA, USA
[8.] Department of Pathology and Laboratory Medicine, Division of Clinical Pathology, The Pennsylvania State University College of Medicine, Hershey, PA, USA
[9.] Department of Veterinary and Biomedical Sciences, The Pennsylvania State University, University Park, PA, USA

**Corresponding Author:** Sharon Xiaolei Huang; 413G Eric. J. Barron Innovation Hub, 123 S. Burrowes Street, State College, PA 16801; (814) 863-7235; suh972@psu.edu

**Author Contributions:** Raman Spectroscopy data analysis, ML algorithm design and implementation, classification and correlation tasks, initial manuscript drafting and supplementary materials preparation were conducted by J. Ye, assisted by Z. Wang and K. Zhang; Y.-T. Yeh, R. Ricker, A. Roder, N. Perea Lopez, H. Liu, N. Zhang and Z. Yu contributed to Raman spectra data collection, manuscript drafting and figure design. H. Lu, S. Hafenstein, L. Organtini, W. Green, and E. Ghedin contributed virus samples and helped with paper writing; Y.-T. Yeh, Y. Xue, E. Ghedin, and M. Terrones, S. Huang, and S.X. Huang conceived the initial ideas and helped with paper writing.

**This file includes:**

Main Text
References
Figure Legends

## Abstract

Rapid identification of newly emerging or circulating viruses is an important first step toward managing the public health response to potential outbreaks. A portable virus capture device coupled with label-free Raman Spectroscopy holds the promise of fast detection by rapidly obtaining the Raman signature of a virus followed by a machine learning approach applied to recognize the virus based on its Raman spectrum, which is used as a fingerprint. We present such a machine learning approach for analyzing Raman spectra of human and avian viruses. A Convolutional Neural Network (CNN) classifier specifically designed for spectral data achieves very high accuracy for a variety of virus type or subtype identification tasks. In particular, it achieves 99% accuracy for classifying influenza virus type A vs. type B, 96% accuracy for classifying four subtypes of influenza A, 95% accuracy for differentiating enveloped and non-enveloped viruses, and 99% accuracy for differentiating avian coronavirus (infectious bronchitis virus, IBV) from other avian viruses. Furthermore, interpretation of neural net responses in the trained CNN model using a full-gradient algorithm highlights Raman spectral ranges that are most important to virus identification. By correlating ML-selected salient Raman ranges with the signature ranges of known biomolecules and chemical functional groups—for example, amide, amino acid, carboxylic acid—we verify that our ML model effectively recognizes the Raman signatures of proteins, lipids and other vital functional groups present in different viruses and uses a weighted combination of these signatures to identify viruses.

## Significance Statement

A large Raman dataset collected on a variety of viruses enables the training of machine learning (ML) models capable of highly accurate and sensitive virus identification. The trained ML models can then be integrated with a portable device to provide real-time virus detection and identification capability. We validate this conceptual framework by presenting highly accurate virus type and subtype identification results using a convolutional neural network to classify Raman spectra of viruses. The accurate and interpretable machine learning model developed for Raman virus identification presents promising potential in a real-time label-free virus detection system that could be used in future outbreaks and pandemics.

## Introduction

Viral outbreaks can spread very quickly through various populations and lead to epidemics, and in some cases, pandemics. Seasonal influenza takes an estimated 389,000 lives globally each year (1), and the SARS-CoV-2 pandemic that began in late 2019 has caused more than 167 million infections and over 3.46 million reported deaths globally (2). These infections also come at a tremendous cost to the global economy and threaten to overwhelm healthcare systems. Therefore, it is critically essential to predict, monitor and control virus infection outbreaks by timely and accurately identifying emerging virus strains.

In the case of an outbreak, rapid identification and detection is often the first step for an effective public health response (3). Once a pathogen has been identified, polymerase chain reaction (PCR) diagnostic testing is often the gold standard to detect viruses as it provides high sensitivity and high specificity. However, the turnaround time, often of several hours, and the fact that it requires targeted detection, makes it a limited approach for a rapid response. Rapid tests based on antigen detection have a quick turnaround time of a few minutes, but sensitivity is often low. The ideal set-up for rapid diagnostics as well as early detection of new circulating virus types, subtypes, or antigenic variants to inform surveillance and vaccine development is a platform that employs little pre-processing of the samples and has fast, unbiased, and sensitive detection capabilities.

A handheld device that could be taken into the field or clinics would be extremely powerful and quickly become the standard approach for virus surveillance. The prototype of such a portable device was previously proposed (4), known as VIRRION, based on a micro-fluidic platform containing carbon nanotube (CNT) arrays for label-free capture and enrichment of viruses from clinical samples coupled with an optical detection technology using surface-enhanced Raman spectroscopy that is sensitive to surface proteins and other components of viruses. The input to such a device can be virus cultures, saliva, nasal washes, or even exhaled breath. The output of the device is the Raman spectra of captured viruses. Combining the device with advanced machine learning models that can classify these spectra to identify the type, subtype, and strain of captured viruses promises an innovative system which can quickly detect, track and monitor viral outbreaks in real-time.

Machine learning (ML) has been successfully applied to Raman spectroscopy analysis in various application scenarios, such as cancer detection (5) and bacteria classification (6). One limitation of current ML-based spectra analysis methods is the lack of transparency in the decision-making process and lack of interpretation of the ML models. Although high accuracy is often reported, the trained ML models, especially those based on deep learning (1), are not transparent and do not provide insight into why and how such accuracy is achieved. One approach to enhance transparency is to develop ML models that highlight salient features used for virus identification and then correlate such ML-selected features (e.g. Raman wavenumber ranges) with the Raman signatures of biomolecules known to exist in viruses such as proteins and lipids (7, 8). A previous study has shown that influenza viruses can be identified by Raman signals generated by surface proteins and lipids (9). Another study on SARS-CoV-2 detected peaks corresponding to the spike protein using Raman spectroscopy (10). However, these existing studies lack quantitative analyses and peak-matching to functional groups (11–13).

In this work, we aim to develop a highly accurate and interpretable machine learning framework for virus identification based on Raman spectra. We propose a one-dimensional Convolutional Neural Network (CNN) that is specifically designed to extract multi-scale features from 1D Raman spectra and perform classification based on the extracted features. Compared to existing ML models, our 1D CNN model is made more interpretable for Raman spectra analysis by incorporating a full-gradient algorithm that calculates a "feature importance map", which shows the relative importance of wavenumbers in recognizing the corresponding virus types of input spectra. When tested on our dataset of virus Raman spectra, the CNN model achieves at least 95% accuracy for classifying different types of Influenza virus and different subtypes of Influenza

A virus, differentiating enveloped from non-enveloped viruses, and differentiating avian coronavirus (infectious bronchitis virus, IBV) from other avian viruses. The wavenumbers (or Raman features) highlighted by the CNN-calculated feature importance map can also inform us on what virus features Raman spectroscopy detects and that machine learning employs for identification. To better understand the molecular basis of Raman detection, we gathered information from the literature on Raman signatures of protein-related functional groups such as amide, amino acid, carboxylic acid; lipids and lipid-related functional groups such as aliphatic chains (7, 8, 14), and collected data in our own set of experiments using protein domains of interest. We find that these known signature ranges correlate well with the key Raman frequency ranges located by our CNN-based ML model. We also designed a novel quantifiable metric to measure the level of correlation between the ML-selected ranges and the signature ranges of specific biomolecules.

## Results

A schematic demonstration of the VIRRION platform (4) for label-free capture and enrichment of viruses is shown in **Fig. 1*A***. We used the device to acquire a dataset consisting of Raman spectra of three groups of RNA viruses, including human respiratory viruses (influenza A H1N1 and H3N2, influenza B, rhinovirus, respiratory syncytial virus (RSV)), avian respiratory viruses (influenza A H5N2 and H7N2, infectious bronchitis virus (IBV), reovirus), and human enteroviruses (Coxsackievirus B type 1 and 3 (CVB1, CVB3), enteroviruses EV70 and EV71). Details of the virus sample preparation procedures can be found in the *Virus samples preparation* subsection under the *Materials and Methods* section. Machine learning experiments were then conducted using this dataset of Raman spectra of viruses. **Fig. 1*B*** shows the architecture of our proposed one-dimensional convolutional neural network (1D-CNN) for classification of virus spectra and illustrates the idea that our machine learning framework can be applied to interpret Raman wavenumber ranges important to ML classification with respect to their correlation with Raman peak ranges of various biomolecules existing in viruses.

**Virus Raman Spectra Data Pre-processing and Augmentation**

Before feeding the Raman spectra into ML classifiers as input, it is essential to employ a few preprocessing steps to reduce noise in the spectra that could potentially undermine the classification performance of trained ML models. One important preprocessing step is baseline correction. We applied the asymmetric least squares smoothing algorithm (15) for baseline correction on each Raman spectrum for all types and subtypes of viruses. In **Fig. 2*A***, we show human influenza A (FLUA) and influenza B (FLUB) example spectra before and after baseline correction. For illustration of the spectrum data distribution after baseline correction, we visualize the FLUA and FLUB spectra using a t-distributed stochastic neighbor embedding (t-SNE) (16) plot (Fig. 2*B*). In *SI Appendix*, **Fig. S1** we further compare the t-SNE plots before and after baseline correction, for all spectra of all virus types in our entire dataset. From the comparison, we observe that applying baseline correction makes the spectra of different viruses more distinguishable, which makes it easier to achieve high accuracy in virus classification tasks. More details about the baseline correction algorithm and parameters used for generating the t-SNE plots are explained in the *Virus Raman spectra preprocessing* subsection under the *Materials and Methods* section.

In **Fig. 3** we show the number of Raman spectra for the human respiratory viruses, avian viruses, and human enteroviruses in our dataset. Considering that the number of spectra vary among viruses (indicating the presence of data imbalance), we adopted a data augmentation strategy by random oversampling (17). For any classification task, the oversampling augmentation is implemented for virus types with fewer spectra in the training set by bootstrapping, a statistical technique that samples data with replacement (18), so that after the augmentation, the number of spectra of every virus type matches that of the virus type with the largest number of training spectra for the task.

**Convolutional Neural Network for Classification of Virus Raman Spectra**

To perform virus identification from Raman spectra, we compared the performances of several different machine learning models including XGBoost (19), and CNN. XGBoost is a popular ML method similar to the Random Forest (20) method. Instead of an ensemble of multiple decision trees in a random forest, XGBoost uses a boosting style ensemble which iteratively builds more decision trees in the learning process. CNN, in comparison, has stronger capability in learning feature representations. However, widely used 2D convolutional kernels are not appropriate for sequence-like data such as Raman spectra. To this end, we designed a one-dimensional CNN to extract features from Raman spectra and perform accurate virus identification. **Figure 1*B*** demonstrates the architecture of our proposed 1D-CNN for virus classification using spectra. Inputs to both 1D-CNN and XGBoost are Raman spectra in the format of 1D vectors. Details about the architecture and training process of our CNN classifier can be found in the *CNN architecture and training details* subsection under the *Materials and Methods* section. For experiments using XGBoost, we kept the built-in default setting of XGBoost (19).

We measured classification performance using three metrics (accuracy, sensitivity, and specificity). The mathematical definitions for these metrics are provided in **Table S1** in *SI Appendix*. Comparing CNN and XGBoost, our 1D-CNN model achieved better performance in all classification tasks, including virus identification from all possible virus types, differentiating enveloped from non-enveloped viruses, classifying different types of human respiratory viruses, differentiating human FLUA from FLUB viruses, identifying the subtype of FLUA viruses, and classifying avian viruses; **Fig. 4** summarizes the classification results. The actual metric numbers for each virus group and all classification experiments are included in *SI Appendix*, **Fig. S2-S7**. Among all virus types and subtypes, the CNN classifier achieved the highest identification accuracies for IBV coronavirus and FLUA virus, around 98% and 97%, respectively (*SI Appendix*, **Fig. S8**).

**Interpretation of Salient Raman Ranges Selected by Machine Learning**

While CNN achieves promising classification results, neural networks including CNNs are known to be "black boxes" and often do not provide sufficient explanation for the learned feature representations (21). Recent advances in interpretability of Neural Network models have alleviated this concern by offering numerous ways of visualizing the weights and features within the NN layers (22–27). Here we propose a method of interpreting our 1D-CNN decision-making process by calculating a "feature importance map", which shows the relative importance of wavenumbers in recognizing the corresponding virus types of input spectra. The wavenumbers (or features) with the highest importance values can tell us what virus features Raman spectroscopy detects that machine learning uses to identify the viruses. The calculation of the feature importance map is based on a full-gradients algorithm (28), as illustrated in the overview diagram **Fig. 1*B*** and detailed in the *Calculation of Raman feature importance maps using CNN neural network responses* subsection under the *Materials and Methods* section.

The feature importance map allows us to identify Raman signature ranges deemed most important by the CNN classifier for virus identification. We can then correlate these ML-selected salient Raman ranges with the signature peak ranges of known biomolecules and chemical functional groups such as lipids, proteins, nucleic acids, amino acids and amide, to seek insights into what differences in biomolecular composition among viruses are captured in the Raman spectra and then used by ML to recognize viruses.

To measure the level of correlation between the ML-selected important wavenumber ranges and Raman peak wavenumber ranges of a known biomolecule, we propose a quantifiable metric termed "matching score". It is a ratio with the numerator as the range of overlapped wavenumbers between ML-determined important ranges and Raman peak ranges of the biomolecule, and the denominator as the total Raman peak ranges of that biomolecule (**Fig. 5**). The higher the matching score, the more likely the signatures of the biomolecule contribute substantially to distinguish viruses. Using this quantifiable metric, we can make some educated guesses about the relative importance of biomolecules in virus identification tasks. Details of this novel algorithm for measuring correlation can be found in the *Interpretable Raman Signatures* subsection under the *Materials and Methods* section*.*

In **Figs. 6-8** we show example feature importance maps from CNN and their correlation with Raman peak ranges of biomolecules known to exist in viruses. In choosing which biomolecules and functional groups to evaluate, we used prior knowledge about the composition of the RNA viruses in our study. Some viruses are enveloped (FLUA and FLUB, IBV coronavirus, RSV), some are not (Reovirus, Enterovirus CVB1/CVB3/EV70/EV71/PV2, Rhino), thus, we included lipid as one type of biomolecule to evaluate since the envelope is formed by the cell surface lipid bilayers. We also included surface protein-related functional groups and individual amino acids, such as amide, phenylalanine and tyrosine. Phenylalanine and tyrosine are chosen because of reports that they are present and important in respiratory viruses (29–33). RNA is also included because all viruses tested here have RNA genomes. Details about how we obtained the Raman peak ranges for the biomolecules and functional groups under consideration are available in the *Interpretable Raman signatures* subsection under the *Materials and Methods* section. Next, we calculated the matching scores between ML-selected important Raman ranges and biomolecule peak ranges for various virus classification tasks.

*Enveloped vs. Non-enveloped Virus Classification.* We did an experiment to train an ML model to classify enveloped vs. non-enveloped viruses and achieved very high accuracy (94.8% accuracy; see *SI Appendix*, **Fig. S6**). This ML model could be used for fast screening, to identify whether a new virus is enveloped or non-enveloped. In **Fig. 6**, we show the Raman feature importance map calculated by this ML model as well as matching scores between ML-selected important ranges and biomolecular peak ranges. From the matching score table, one can see that for this task lipid is shown to be much more important (matching score 51.98%) than protein-related functional groups (matching scores 25% and 7.98% for Amide I and Amide III, respectively). This is consistent with the difference between enveloped and non-enveloped viruses, which is that enveloped viruses have an enclosing phospholipid bilayer whereas non-enveloped viruses do not have the phospholipid bilayer. It is highly likely that the ML model is picking up the signature ranges of lipid to differentiate enveloped from non-enveloped viruses.

*Comparison of Classification Tasks that Differentiate Various Flu Types and Subtypes.* We trained several ML models to differentiate influenza viruses such as avian FLUA from human FLUA, and human FLUA from human FLUB (**Fig. 7** and *SI Appendix,* **Fig. S5**). From the matching score table, we noted that the Amide III range is not important for classifying avian FLUA vs. human FLUA (matching score 13.16%) but more important when differentiating human FLUA from human FLUB (matching score 73.68%). Lipid is more important when differentiating avian FLUA from human FLUA but less important when differentiating human FLUA from human FLUB, likely indicating that the ML model trained for classifying avian FLUA vs. human FLUA is capturing their differences in the envelopes since the phospholipid bilayer of the human viruses comes from different cells than the avian viruses that were isolated in eggs. Also, the RNA matching score stands out to be higher (60%) when differentiating human FLUA from human FLUB, compared to classifying avian FLUA vs. human FLUA (40%). In another experiment, we trained a machine learning model to differentiate four subtypes of FLUA, human H1N1, H3N2, and avian H5N2 and H7N2 (See *SI Appendix*, **Fig. S4**). Again, we observe that lipid is important whereas Amide III is not important when differentiating the FLUA subtypes.

*Classification of Avian Viruses including IBV Coronavirus.* We trained an ML model to differentiate three types of avian viruses and achieved very high accuracy (99.8%) in identifying the IBV coronavirus (**Fig. 8**, and *SI Appendix*, **Fig. S2**). This shows that the Raman spectra of coronavirus have specific signatures that make them easily identifiable when compared to avian influenza. The proposed technique combining Raman spectroscopy and ML could potentially be used for highly reliable detection and identification of coronaviruses. From the matching score table shown in **Fig.**

**8**, one can see that both lipid and protein peak ranges have high correlation with ML-selected important Raman ranges for distinguishing IBV coronavirus from other avian viruses, likely indicating that Raman spectroscopy and ML are picking up signatures of the spike protein and receptor binding domains of coronaviruses.

*Additional Observations about Correlation between ML-selected Important Raman Ranges and Biomolecule Peak Ranges.* When comparing the matching scores for the experiment classifying different human respiratory viruses (See *SI Appendix*, **Fig. S7**) and the experiment classifying different subtypes of Influenza (FLU) A (See *SI Appendix*, **Fig. S4**), we observe that: 1) the relative importance of lipids is higher in the FLUA subtype identification task. 2) There is a significant difference in the relative importance of the Amide III range. While Amide III is very important in respiratory virus classification, it is minimally important in FLUA subtype identification, which could indicate that the spectra of all subtypes of FLUA are very similar in the Amide III range. Amide III is a signature Raman band in proteins but can be sensitive to secondary structures, and such a difference of its matching scores indicates that the viruses have different surface proteins (34, 35) —which is indeed the case when comparing different families of viruses—and that the subtypes have slight differences in their surface proteins—which again is the case since FLUA viruses have Hemagglutinin (HA) and Neuraminidase (NA) on their surface. This is consistent with our preliminary findings about the viruses under study. 3) The two chosen amino acids (phenylalanine and tyrosine) are consistently important in respiratory virus type or subtype classification. And, finally, 4) the RNA genome is also generally important for detecting virus differences by Raman.

**Viral Dose Detection Limit.**

To determine what the viral dose detection limit of our method was, we conducted a series of dilution experiments using a flu virus dataset consisting of Raman spectra of 11 influenza virus strains. Information about this dataset is given in **Table S2** in *SI Appendix.* Around 10,000 spectra were collected for each virus sample at the original undiluted concentration. For two strains– A/Indiana/08/2018(H3N2) and A/Nebraska/14/2019(H1N1)—we collected spectra at different virus concentrations. The original undiluted viruses were for Indiana/08 at a TCID50 (Median Tissue Culture Infectious Dose) of 1.45e+07 viruses/ml and an RNA copy number of 1.42e+09/ml; while Nebraska/14 was at a TCID50 of 2.29e+07/ml and an RNA copy number of 2.27e+09/ml. Increments of 10-fold dilutions were performed down to $10^{-6}$ dilution, which corresponded to fewer than one replicating virus and approximately 14 RNA copies per 10μL of solution for Indiana/08, and less than one replicating virus and approximately 23 RNA copies per 10μL of solution for Nebraska/14. **Table S3** in *SI Appendix* displays the expected number of viruses and RNA copies in each 10 μL of sample solution used to collect spectra. At each level of dilution, 400 Raman spectra were collected for the corresponding 10 μL of solution.

On this dataset, we conducted ML experiments using our proposed CNN model as shown in **Figure 1B**, in a blind-testing setting, for flu type and subtype classification. Since the two strains being used for testing the viral dose detection limit, Indiana/08 and Nebraska/14, are among the 11 strains in the dataset, we trained our ML model using 9 strains of H1N1, H3N2, and FluB in **Table S2** (see *SI Appendix*), excluding these two testing strains. Then, the spectra of the two testing strains at different dilution levels were classified using the trained ML model, as previously unseen strains (i.e. not contained in the training set). The goal was to examine the ML classification performance for spectra collected at different dilutions and thus infer the detection limit of our approach. The ML classification results are shown in **Table S4** in *SI Appendix.* We can observe from the results that our ML model, which was trained on 9 strains (not including the two testing strains) using spectra collected at the undiluted concentration only, was able to reliably predict the subtype of the two testing strains using spectra collected at the undiluted, and $10^{-1}$, $10^{-2}$, $10^{-3}$, $10^{-4}$, $10^{-5}$ dilutions. At $10^{-6}$ dilution, however, we start to see unpredictability; in the case of Indiana/08, the percentage of blank spectra among the 400 collected spectra spiked to 90.25%, which means that we had to filter out around 90% of the spectra in order to correctly classify the case, and for Nebraska/14, our model mistakenly classified it to Flu B using spectra collected at $10^{-6}$ dilution. Therefore, based on this set of experiments, the detection limit of our technique in the present set up is $10^{-5}$ dilution, which corresponds to roughly 1 replicating virus and 142 RNA copies per 10μL for Indiana/08, and roughly 2 replicating viruses and 227 RNA copies per 10μL for Nebraska/14 (see Table S3 in *SI Appendix*).

## Discussion

Detecting and classifying virus using the technique presented here is very fast, making it feasible as a real-time, label-free virus screening and detection tool. Once the ML model is trained, it takes approximately 1e-05s on an NVIDIA Quadro RTX 6000 GPU with 24GB RAM. If we are performing case-based experiments, i.e. classifying hundreds of spectra collected for a virus sample to determine the virus label, the run time is still less than a second. The more challenging aspect is whether the detection can extend to viruses not contained in the training set. The blind-testing experiments conducted for determining the viral dose detection limit (see **Table S4** in *SI Appendix*) demonstrate that while the ML model can recognize the type and subtype of a virus not contained in the training set, it may not be able to recognize the specific strain (often determined by the year and region the virus was isolated). The model can still predict the broader category (type, subtype) of a strain in the training set that is recognized as closest to the unseen strain because of the model's ability to output a probability score and correlate the Raman signature of the testing strain with those of known strains. Being able to detect an unknown strain and interpret its Raman signature is one of our main future research directions. We are investigating zero-shot machine learning techniques that can be integrated into our ML framework so that our model will be able to either detect a virus strain that is already contained in the training set or predict that the testing virus is of a previously unknown strain and in such case interpret its Raman signature in terms of its correlation with the Raman signatures of known viruses, biomolecules, and/or chemical functional groups present in viruses. The expected outcome of such an improved model is that the model will first provide a binary decision regarding whether the testing strain is one of the strains in the training set (i.e. same strain, but different samples). If the strain is recognized as one of those seen ones, its label will be predicted. If the strain is detected as a new strain that is not contained in the training set, the feature importance map output by our model will allow us to examine where (i.e., based on which biomolecules or chemical functional groups) the new strain is different from previously seen strains. The implication is that the model could then predict which existing strains are the closest to the new strain.

The robustness of a virus screening and detection system using our technique can be improved through more robust spectra collection, and by refining pre-processing steps to ensure the quality of the spectra used in the ML experiments. A practical system will consist of virus capture and enrichment, spectra collection, and a sequence of pre-processing steps to prune out outlier spectra and remove blank spectra. The remaining Raman spectra encoding virus signatures are then classified by the trained ML model to recognize the virus label. One encouraging observation is that adding a pre-processing step to remove blank spectra has been extremely helpful in improving classification performance (see **Table S4** in *SI Appendix*). Note that the blank spectra were identified by a blank-spectra classifier which was trained using a small dataset of blank spectra collected from only background and no virus. With pre-processing steps such as discarding blank spectra and possibly other outlier spectra with signal from a more "real-life" background media, our technique will be more robust because it can then filter out spectra that do not encode virus signal. The potential of using pre-processing to remove noisy and irrelevant spectra could also explain one phenomenon that we observe from Table S4 in *SI Appendix*, which is that the accuracies at more diluted levels from $10^{-1}$ to $10^{-5}$ do not decrease and can still be high, maybe because contaminants and background are also being removed at higher dilution. Thus the remaining spectra being classified by the ML model are "cleaner" virus spectra. This may not be the case in clinical samples with a low concentration of virus as background molecules, in this case, would not be diluted. However, host contamination (background molecules) would in principle get filtered out when run through the carbon nanotubes (see (4)). Furthermore, pre-processing and filtering steps can be applied to remove spectra resulting from background and leave only spectra with virus signal to be classified by the ML model. While determining how the system reacts with real biological fluids and tissues (saliva, etc.) is of very high interest to us, this represents the next step and goes beyond the scope of our current study.

While the methods we present here are for rapid detection, it is not meant to replace Polymerase chain reaction (PCR), which is a highly specific and sensitive method for the detection of known viruses (36). Our goal was to develop a ML approach to better mine the spectra from Raman spectroscopy for rapid and label-free detection of viruses. In the process we also present important findings about Raman signatures of virus-related biomolecules that are utilized by the interpretable machine learning model for recognizing viruses. Testing the system using saliva and other clinical specimens will require an extensive study to determine how to control for background in various tissues. We will validate the method and the microfluidics device further in our future work.

## Conclusion

In summary, we applied ML to identify viruses imaged by Raman spectroscopy. Our ML system based on a CNN implementation shows high accuracy in classifying different types of human and avian viruses. It can also differentiate subtypes of influenza A viruses. The interpretation of the neural network responses also provides valuable information about Raman wavenumber ranges that correlate well with the signature ranges of known biomolecules and chemical functional groups present in viruses. The major contributions of our work are that:

(1) We developed a 1D-CNN classifier that achieved high accuracy for multiple virus identification and classification tasks, including differentiating enveloped from non-enveloped viruses, identifying types of human respiratory viruses, differentiating human FLUA from human FLUB viruses, classifying subtypes of Influenza A viruses, and differentiating among types of avian viruses;

(2) We further investigated the association between classifier-selected important Raman ranges and peak ranges of lipids, proteins and relevant chemical functional groups, and observed correlations that are consistent with existing knowledge;

(3) We delivered promising virus classification results that indicate Raman spectra of different virus types and subtypes contain recognizable Raman signatures that can be identified by ML models, which unravels the potential of using interpretable ML in a real-time virus surveillance system.

In our future work, we will collect more Raman spectra of different virus samples (human and animal, including DNA viruses) to build a large virus spectra database for training robust and highly accurate ML models. We will study virus evolution using temporal ML models trained on Raman spectra of virus strains of different types, from different years and locations. And we will further improve the Raman enhancement with better signal intensities and lower noise levels, considering the feedback from ML classification and feature importance identification.

## Materials and Methods

**Virus Samples Preparation.**
Avian influenza virus (AIV) was propagated in specific-pathogen-free (SPF) embryonating chicken eggs (ECE) via allantoic cavity route inoculation at 9-11 days of age. The inoculated ECEs were incubated in a 37°C egg incubator for 3d (or 72 h) and then were removed/chilled at 4 °C for a minimum of 4 h or overnight. Allantoic fluid (AF) containing the virus was harvested from each egg using a sterile technique (a 3 mL sterile syringe with a 25G×5/8” needle). The harvested AF was clarified by centrifugation at 8000-1000 rpm for 10 min. Virus titer was determined in embryo infectious doses 50% ($EID_{50}$) titers by the Reed-Muench method (37). Briefly, the $EID_{50}$ test was conducted in ECE. The propagated fresh stock H5N2 AIV was prepared in 10-fold serial dilutions from $10^{-1}$ through $10^{-9}$. Each dilution was inoculated into 5 eggs, 0.1 mL per egg. The inoculated eggs were incubated at 37 °C for 72 hours. The eggs were candled daily to remove dead eggs to chill them at 4 °C refrigerator. After 72 hours of incubation, allantoic fluid was harvested from each egg (38).

H1N1, H3N2, FluB, Rhinovirus, and RSV were prepared in Madin-Darby canine kidney-London (MDCK-London) cell culture. MDCK-London cells were cultured in Dulbecco's modified Eagle's medium (DMEM; Invitrogen, Carlsbad, CA) containing 10% fetal bovine serum and 1% penicillin-streptomycin, and incubated at 37°C in a humidified $CO_2$ incubator.

Enteroviruses (CVB1, CVB3, EV70, EV71, or PV2) were propagated using HEK 293 cell lines. Infection of cells with enterovirus inoculum, harvesting of cells and media, and additional virus sample preparation steps are documented in the *Virus Propagation and Purification* subsection in (39).

**Virus Raman Spectra Data Acquisition.**
The VIRRION platform (4) constructed with nitrogen-doped carbon nanotube arrays and gold (Au) nanoparticles was used as SERS substrate for collecting Raman spectra from virus samples. A 100 µL sample of each virus was directly dropped (drop-cast) onto the VIRRION Au-CNT substrate and air-dried at room temperature for 10 hrs prior to Raman measurements. Raman data acquisition was recorded using a Horiba-LabRAM HR Evolution system with a 785nm diode laser line. The laser power on the sample was ca. 3.6 mW, focused through a 100× objective. The 600 gr/mm grating was used with a spectral range from 500 $cm^{-1}$ to 2000 $cm^{-1}$. The typical acquisition time was 30 s.

**Virus Raman Spectra Preprocessing Algorithm Details.**
First, we apply baseline correction with asymmetric least squares smoothing (15) to reduce background noise in spectra. This method estimates a polynomial baseline to correct baseline shift in Raman measurements. Then, we adjust the intensities that vary across the spectra of different virus types to a universal scale by normalization. In this step, we apply $L_2$ normalization that converts the input vectors to unit vectors. The normalization makes intensities comparable across spectra and facilitates convergence during ML model training.

For generating the t-SNE plots of spectra data (**Fig. 2** and *SI Appendix*, **Fig. S1**), we use the scikit-learn (40) machine learning package to perform the t-SNE dimensionality reduction and map high-dimensional data points to a two-dimensional space. We use the default parameters of the package except for the perplexity value which we set to 50 and the learning rate to 200. Under this setting of parameters, the data points in the two-dimensional map for FLUA and FLUB spectra fall into clearly distinct clusters, which indicates that a deep learning network capable of nonlinear functional mapping should be able to achieve highly accurate classification on the dataset.

**CNN Architecture and Training Details.**
As shown in Fig. 1*B*, the Convolutional Neural Network for our task is built with 4 convolutional blocks. Considering the dimension of our training set is $\mathrm{N}\times 1\times \mathrm{D_w}$, where $\mathrm{N}$ refers to the number of Raman spectra samples in the training set and $\mathrm{D_w}$ is the dimension of Raman wavenumber range, a reasonable option for the convolutional blocks is to adopt 1D CNN layer. Followed by the convolutional layer is a 1D batch normalization layer and an activation layer, in this case we choose ReLU. The kernel size and stride of the 1D CNN layer of the first convolutional block are both set as 1, with the width fixed while increasing the depth from the input dimension 1 to the dimension of the hidden state, which will be specified later along with other hyper-parameters. Next, for the other three convolutional blocks, kernel size is increased to 3 and stride is set as 2 for reducing the dimension of feature maps by half each time. Followed by the activation layer of the second and the third convolutional blocks, 2 dropout layers with rates 0.5 and 0.25 are applied, respectively, for alleviating over-fitting to the training set. After all convolutional blocks, the last layer for obtaining the final classification results is a fully connected layer with output dimension as $\mathrm{N}\times 1\times \mathrm{D_c}$, where $\mathrm{D_c}$ is denoted as the number of virus types or subtypes, depending on the classification task and specific dataset used for that task.

During training, we apply a 5-fold cross-validation and stratified sampling for each fold based on the virus types (or subtypes) to ensure that after splitting the dataset into training and testing sets, every type (or subtype) gets equal representation in both sets regardless of how unbalanced the data distribution is. For fair comparison, we run the cross-validation 5 times and obtain the average score for all metrics across the 5 test runs. The corresponding performances reported in Figure 4 are averaged results among the 5 hold-out test sets from cross-validation with error bars. Learning curves of the 5-fold cross validation for the classification task on Flu A subtypes (H1N1/H3N2/H5N2/H7N2) are shown in *SI Appendix,* ***Fig. S10*****.** The process of setting the hyperparameters was performed by manually fine-tuning and choosing the hyper-parameters that gave good results, for our 1D-CNN model. All hyper-parameters are fixed for each run, the learning rate is set as 0.001 and trained for 1000 epochs with hidden dimension set to 128 for the first convolution block, and then decreases by half for every subsequent convolutional block. The Adam optimizer is used, and dropout rate is set as 0.2. Although systematic grid search for optimal hyperparameters was not needed in this work because the accuracy levels are relatively high already with the manually-set hyperparameters, we expect that grid search optimization may be needed when we extend our dataset to include more viruses and larger sets of Raman spectra.

**Calculation of Raman Feature Importance Map using CNN Neural Network Responses.**
While the CNN classifiers trained for virus identification tasks achieve high performance, we are interested in learning what Raman features are utilized by these classifiers to differentiate among viruses. To this end, we propose an algorithm for the CNN to infer the feature importance value for each specific wavenumber for further investigation of interpretability. With regards to the interpretation of feature extraction and selection by neural networks, a saliency map has been widely considered as an intuitive and well-established method to visualize the importance value for each unit within the input data (22–27). However, in our case, the contribution each wavenumber has to the final virus type (or subtype) classification is highly unlikely to be independent from each other. A more reasonable assumption is that the distinguishable features from Raman spectra of a specific virus type (or subtype) are composed of a set of Raman signature ranges, besides individual wavenumbers of Raman spectra. Hence a desirable design of saliency map representation for interpretation is expected to include both attributions to ensure the completeness of the feature map. By leveraging features from input vector and neurons from intermediate layers simultaneously, the full-gradient algorithm (28) is proven to be a sensible representation of CNN interpretability in terms of the capability to capture both local and global attributions from each Raman spectra wavenumber and signature ranges. As the full gradient representation for neural network visualization (28) was originally designed and applied on natural images, we adapt and modify the full-gradient algorithm to accommodate the 1D vector inputs, as the format of Raman spectra is in our case. As shown in **Fig. 1*B***, the process for extracting feature importance for each virus type or subtype is demonstrated below the architecture of CNN. The full gradient feature importance map extracted is defined as

$$S_f(x) = \psi(\nabla_x f(x) \odot x) + \sum_{l \in L} \sum_{c \in c_l} \psi(\nabla_b f(x,b)_c \odot b).$$

Here, the saliency map of the full-gradient representation consists of two parts---input gradient that is specific to each wavenumber of Raman spectra in the training set, and bias-gradient from each convolutional block. The components of each convolutional block are illustrated in Fig. 1*B*. The approximate network-wide representation of the feature map is considered comprehensive for capturing what the model learned throughout the process of the classification task from both lower and higher levels of abstraction. $c$ refers to the virus type or subtype, depending on the target for a particular classification task. Gradients specific for each $c$ are extracted separately in order to get insights of Raman frequency significance for different types or subtypes of viruses. This process is implemented by activating the virus type (or subtype) of interest during backpropagation through the entire set of convolutional blocks while obtaining the cross-entropy loss for each $c$. $\psi(\cdot)$ refers to the post-processing steps that can be denoted as $\psi(\cdot) = \mathrm{bilinearUpsample}(\mathrm{normalize}(\mathrm{abs}(\cdot)))$. First, the operation that is applicable for both input and bias gradients is the step of obtaining the absolute value of either positive or negative importance to visualize the significance while neglecting the sign. Next, the absolute values of gradients are normalized to the range of $[0, 1]$ to optimize the visualization by creating proper viewing contrast. And then for the gradients extracted from convolutional blocks with dimension downsized to different scales of hidden states, we facilitate the aggregation of the block-wise feature maps by up-sampling each to the same dimension as the input vector with bilinear interpolation. The feature map extracted for each virus type or subtype is shown in the format of the area chart in **Fig. 1*B***.

**Interpretable Raman Signatures.**
Considering that one of our goals is to make an educated guess as to which biomolecules are more likely to have a significant contribution in differentiating virus types or subtypes, we analyze the correlation between the important Raman ranges from CNN feature importance map and the Raman peak ranges of biomolecules existing in viruses. First, the Raman peaks of biomolecules including lipids, proteins, nucleic acids and protein-related chemical functional groups are gathered from the literature (7, 8, 14) (see *SI Appendix*, **Fig. S9** for detailed peak ranges). We note that for a specific functional group, the specific Raman range can vary when measured in different environments. Here we included all the possible Raman ranges for the generality of our analysis. For RBD proteins and amino acids (Tyrosine, Phenylalanine), we measured their Raman spectra in our own experiments and then located the peaks of the Raman spectra by adopting the python package (41). A shift of 5 wavenumbers is granted to each peak to construct the peak ranges for each biomolecule (i.e. as a range for each peak). Second, given a Raman feature (i.e. wavenumber) importance map calculated using the full-gradient algorithm for CNN, we extract important Raman ranges by applying a threshold on the calculated feature importance values. We apply a Savitzky-Golay filter (42) on the relatively noisy importance values, with the length of the filter window set as 17. Then, a 40-percentile threshold is applied to extract ranges in the feature map that consist of wavenumbers with corresponding importance values above the threshold. Finally, the quantifiable metric—the matching score as demonstrated in Fig. 5—is used to measure the level of correlation between Raman peak ranges of biomolecules and important Raman ranges identified by ML. The matching score metric is developed in the format of a ratio, where the numerator of the ratio is the amount of overlap (i.e. number of overlapped wavenumbers) between the ML-calculated important Raman ranges for identifying a particular virus type and the Raman peak ranges of a certain biomolecule, and the denominator is the total number of wavenumbers in the biomolecule's peak ranges. Thus, the matching scores are in the range of [0,1]: a matching score of 1.0 means that the biomolecule's entire peak ranges are considered important by the CNN classifier for identifying the virus; a matching score of 0 indicates no wavenumber within the biomolecule's peak ranges is considered important by the classifier; and when the matching score value is between 0 and 1, the higher the score, the more likely that the biomolecule is important for identifying that particular type of virus. We report the matching scores for all our ML classification tasks in *SI Appendix,* **Fig. S2-S7**, and show the Raman peak ranges for biomolecules in **Fig. S9**.

**Competing Interest Statement:** The authors declare that they have no competing financial interests.

**Data Availability.** Raman spectra of various viruses from the dataset used in this paper are deposited in Figshare (43) and the source code for the 1D-CNN ML model for virus identification using Raman spectra is available on GitHub (44). More data are available upon request, for research purposes only. Please email mtterrones@gmail.com (M.T.) with a short description about the purpose of usage along with your request for more data.

**Acknowledgments.** We thank the National Science Foundation's Growing Convergence Research Big Idea (under Grant ECCS-1934977) and National Science Foundation's Early-concept Grants for Exploratory Research (under Grant OIA-2030857). This work was supported in part by the Division of Intramural Research of the NIAID/NIH (E.G.), and in part by the Huck Institutes of the Life Sciences of the Pennsylvania State University (Y.-T. Y., M.T., S.H., S.X.H.). We also thank the NSF for Grants DMR-1420620 and DMR-2011839 through the Pennsylvania

State University Materials Research Science and Engineering Center (MRSEC)_Center for Nanoscale Science for partial financial support.

**References**

1. J. Paget, *et al.*, Global mortality associated with seasonal influenza epidemics: New burden estimates and predictors from the GLaMOR Project. *J. Glob. Health* **9**, 020421 (2019).

2. W. H. Organization, Others, Coronavirus disease 2019 (COVID-19): situation report, 82 (2020).

3. F. Keesing, *et al.*, Impacts of biodiversity on the emergence and transmission of infectious diseases. *Nature* **468**, 647–652 (2010).

4. Y.-T. Yeh, *et al.*, A rapid and label-free platform for virus capture and identification from clinical samples. *Proc. Natl. Acad. Sci. U. S. A.* **117**, 895–901 (2020).

5. S. Li, *et al.*, Noninvasive prostate cancer screening based on serum surface-enhanced Raman spectroscopy and support vector machine. *Appl. Phys. Lett.* **105**, 091104 (2014).

6. A. Walter, *et al.*, From bulk to single-cell classification of the filamentous growing Streptomyces bacteria by means of Raman spectroscopy. *Appl. Spectrosc.* **65**, 1116–1125 (2011).

7. K. Czamara, *et al.*, Raman spectroscopy of lipids: a review. *J. Raman Spectrosc.* **46**, 4–20 (2015).

8. D. Němeček, G. J. Thomas Jr, “Raman spectroscopy of viruses and viral proteins” in *Frontiers of Molecular Spectroscopy*, (Elsevier, 2009), pp. 553–595.

9. J.-Y. Lim, *et al.*, Identification of newly emerging influenza viruses by detecting the virally infected cells based on surface enhanced Raman spectroscopy and principal component analysis. *Anal. Chem.* **91**, 5677–5684 (2019).

10. D. Zhang, *et al.*, Ultra-fast and onsite interrogation of Severe Acute Respiratory Syndrome Coronavirus 2 (SARS-CoV-2) in environmental specimens via surface enhanced Raman scattering (SERS). *bioRxiv* (2020) https:/doi.org/10.1101/2020.05.02.20086876.

11. Y. Liu, *et al.*, Label and label-free based surface-enhanced Raman scattering for pathogen bacteria detection: A review. *Biosens. Bioelectron.* **94**, 131–140 (2017).

12. M. Reyes, *et al.*, Exploiting the anti-aggregation of gold nanostars for rapid detection of hand, foot, and mouth disease causing Enterovirus 71 using surface-enhanced Raman spectroscopy. *Anal. Chem.* **89**, 5373–5381 (2017).

13. K. Moor, *et al.*, Noninvasive and label-free determination of virus infected cells by Raman spectroscopy. *J. Biomed. Opt.* **19**, 067003 (2014).

14. Y. H. Ong, M. Lim, Q. Liu, Comparison of principal component analysis and biochemical component analysis in Raman spectroscopy for the discrimination of apoptosis and necrosis in K562 leukemia cells. *Opt. Express* **20**, 22158–22171 (2012).

15. S.-J. Baek, A. Park, Y.-J. Ahn, J. Choo, Baseline correction using asymmetrically reweighted penalized least squares smoothing. *Analyst* **140**, 250–257 (2015).

16. L. van der Maaten, Visualizing Data using t-SNE (2008) (May 23, 2021).

17. C. Shorten, T. M. Khoshgoftaar, A survey on image data augmentation for deep learning. *J. Big Data* **6** (2019).

18. G. James, D. Witten, T. Hastie, R. Tibshirani, *An Introduction to Statistical Learning: with Applications in R* (Springer, New York, NY, 2013).

19. T. Chen, C. Guestrin, XGBoost. In *Proceedings of the 22nd ACM SIGKDD International Conference on Knowledge Discovery and Data Mining - KDD '16*, (ACM Press, 2016) https:/doi.org/10.1145/2939672.2939785.

20. L. Breiman, Random Forests. *Mach. Learn.* **45**, 5–32 (2001).

21. J. M. Benitez, J. L. Castro, I. Requena, Are artificial neural networks black boxes? *IEEE Trans. Neural Netw.* **8**, 1156–1164 (1997).

22. M. D. Zeiler, R. Fergus, "Visualizing and understanding convolutional networks". In *Computer Vision – ECCV 2014*, Lecture notes in computer science., (Springer International Publishing, 2014), pp. 818–833.

23. K. Simonyan, A. Vedaldi, A. Zisserman, Deep inside Convolutional Networks: Visualising image classification models and saliency maps. *arXiv [cs.CV]* (2013).

24. J. Yosinski, J. Clune, A. Nguyen, T. Fuchs, H. Lipson, Understanding neural networks through deep visualization. *arXiv [cs.CV]* (2015).

25. R. R. Selvaraju, *et al.*, Grad-CAM: Visual explanations from deep networks via gradient-based localization. In *2017 IEEE International Conference on Computer Vision (ICCV)*, (IEEE, 2017) https:/doi.org/10.1109/iccv.2017.74.

26. P.-J. Kindermans, *et al.*, "The (Un)reliability of saliency methods". In *Explainable AI: Interpreting, Explaining and Visualizing Deep Learning*, Lecture notes in computer science., (Springer International Publishing, 2019), pp. 267–280.

27. K. Zolna, K. J. Geras, K. Cho, Classifier-agnostic saliency map extraction. *Comput. Vis. Image Underst.* **196**, 102969 (2020).

28. S. Srinivas, F. Fleuret, Full-Gradient Representation for Neural Network Visualization. In *NeurIPS*, (2019).

29. C. E. McBride, C. E. Machamer, A single tyrosine in the severe acute respiratory syndrome coronavirus membrane protein cytoplasmic tail is important for efficient interaction with spike protein. *J. Virol.* **84**, 1891–1901 (2010).

30. S. Wang, *et al.*, AXL is a candidate receptor for SARS-CoV-2 that promotes infection of pulmonary and bronchial epithelial cells. *Cell Res.* **31**, 126–140 (2021).

31. F. Y. Shaikh, *et al.*, A critical phenylalanine residue in the respiratory syncytial virus fusion protein cytoplasmic tail mediates assembly of internal viral proteins into viral filaments and particles. *MBio* **3** (2012).

32. C. J. Stewart, *et al.*, Respiratory syncytial virus and Rhinovirus bronchiolitis are associated with distinct metabolic pathways. *J. Infect. Dis.* **217**, 1160–1169 (2018).

33. J. Xu, *et al.*, Increased mortality of acute respiratory distress syndrome was associated with high levels of plasma phenylalanine. *Respir. Res.* **21**, 99 (2020).

34. R. W. Williams, Protein secondary structure analysis using Raman amide I and amide III spectra. *Methods Enzymol.* **130**, 311–331 (1986).

35. A. Rygula, *et al.*, Raman spectroscopy of proteins: a review. *J. Raman Spectrosc.* **44**, 1061–1076 (2013).

36. K. K.-W. To, *et al.*, Temporal profiles of viral load in posterior oropharyngeal saliva samples and serum antibody responses during infection by SARS-CoV-2: an observational cohort study. *Lancet Infect. Dis.* **20**, 565–574 (2020).

37. G. J. Ebrahim, Virology: principles and applications. J. Carter, V. Saunders (eds). *J. Trop. Pediatr.* **55**, 66–66 (2007).

38. S. Zheng, *et al.*, Sizable tunable enrichment platform for capturing nano particles in a fluid. *US Patent* (2020) (May 30, 2021).

39. K. L. Shingler, *et al.*, The enterovirus 71 A-particle forms a gateway to allow genome release: a cryoEM study of picornavirus uncoating. *PLoS Pathog.* **9**, e10032 https://soundcloud.com/bangtan/sothatiloveyou?in=jiarong-ye/sets/bts-english-songs/s-vQqKYnTwhfl&utm_source=clipboard&utm_medium=text&utm_campaign=social_sharing (2013).

40. F. Pedregosa, *et al.*, Scikit-learn: Machine learning in Python. *Journal of machine Learning research* **12**, 2825–2830 (2011).

41. P. Virtanen, *et al.*, SciPy 1.0: fundamental algorithms for scientific computing in Python. *Nat. Methods* **17**, 261–272 (2020).

42. William H. Press, S. A. Teukolsky, Savitzky-Golay Smoothing Filters. *Comput. Phys.* **4**, 669 (1990).

43. J. Ye, Y.-T. Yeh, Dataset for "Accurate Virus Identification with Interpretable Raman Signatures by Machine Learning." Figshare. https://figshare.com/articles/dataset/pnas_dataset_csv/19426739. Deposited 15 April 2022.

44. J. Ye, Accurate Virus Identification with Interpretable Raman Signatures by Machine Learning. GitHub. https://github.com/karenyyy/Accurate-Virus-Identification-with-Interpretable-Raman-Signatures-by-Machine-Learning-. Deposited 15 April 2022

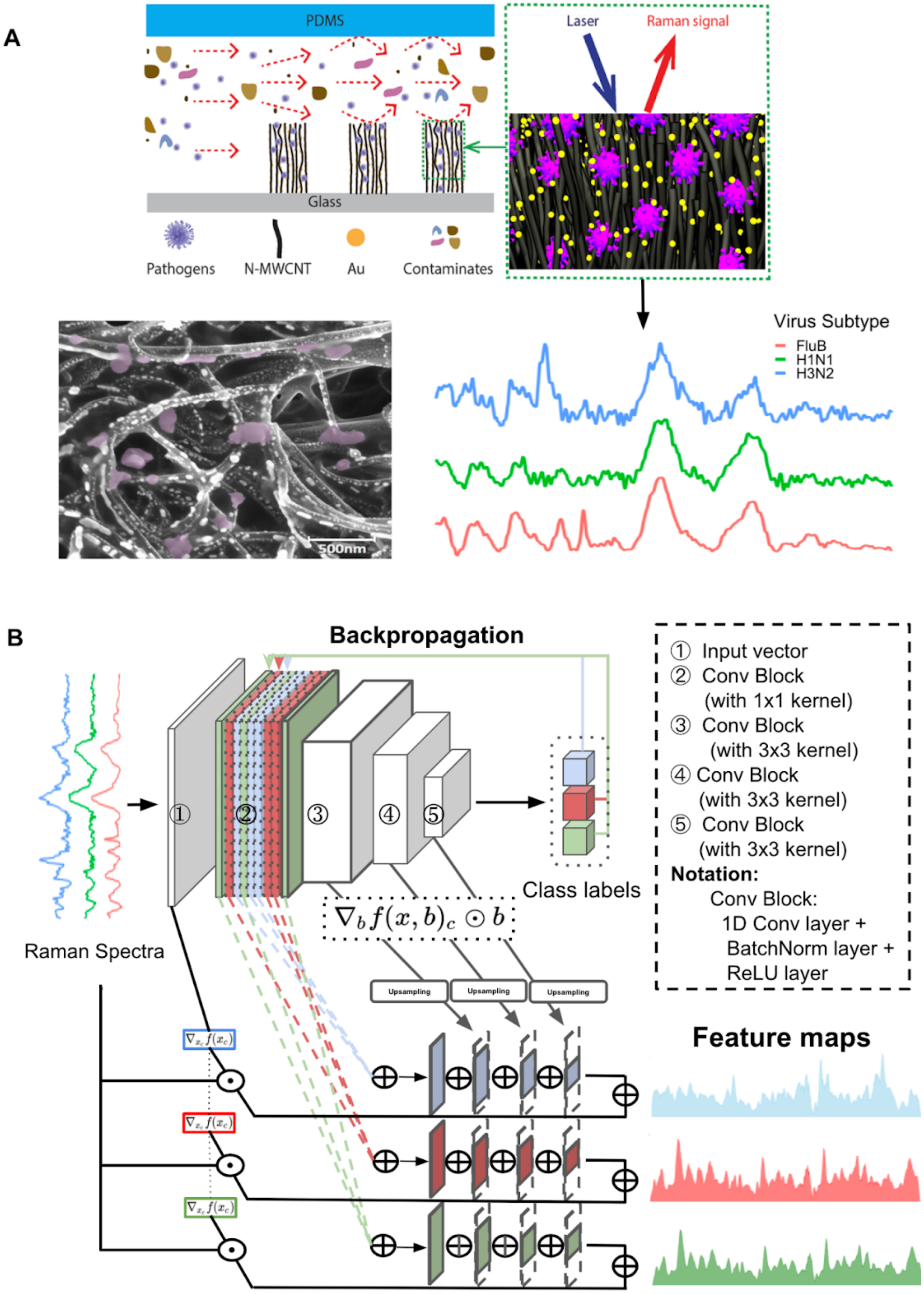

**Figure 1. A.** Schematics showing the nitrogen-doped multiwall carbon nanotubes (N-MWCNTs) device encapsulated in PDMS (polydimethylsiloxane) used to enrich viruses (top left). The viruses are enriched between CNTs where the Au nanoparticles are pre-deposited. Raman spectra are then collected from the virus-enriched samples (top right). A Scanning Electron Microscope (SEM) image (bottom left) of a sample shows CNTs, Au nanoparticles, and trapped viruses (purple colored). Raman spectra from different virus samples are shown (bottom right; FLUB in red, FLUA H1N1 in green and FLUA H3N2 in blue). **B.** The Convolutional Neural Network (CNN) architecture for virus identification and the process of extracting Raman feature maps that show important Raman signature ranges. The feature maps extracted are class-specific, demonstrating the significant Raman ranges for identifying different virus types (or subtypes, depending on the classification task) in different colors.

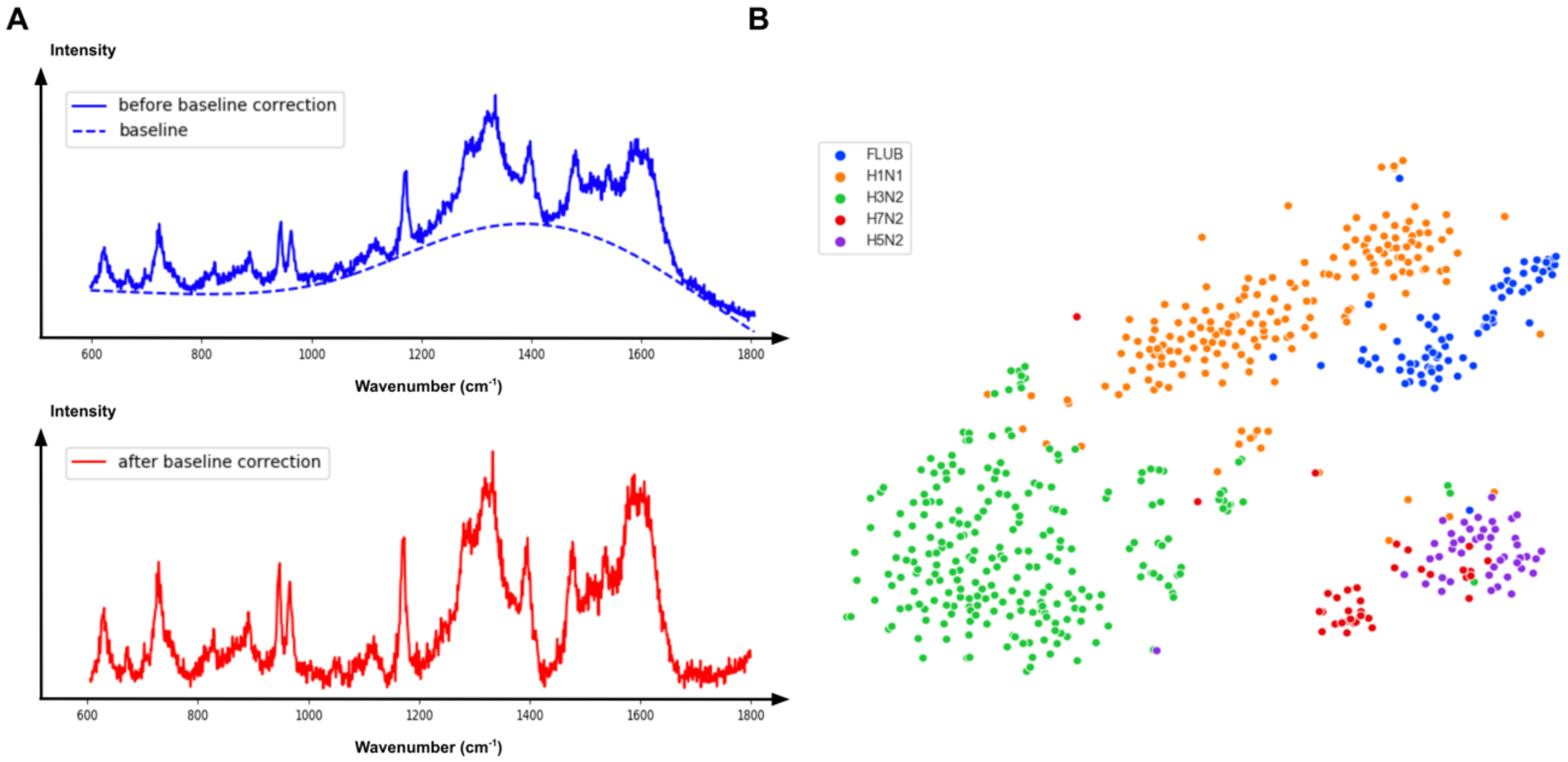

**Figure 2. A.** Sample Raman spectra before and after baseline correction; **B.** T-SNE plot of FLUA subtypes (H1N1, H3N2, H5N2, H7N2) and FLUB after baseline correction.

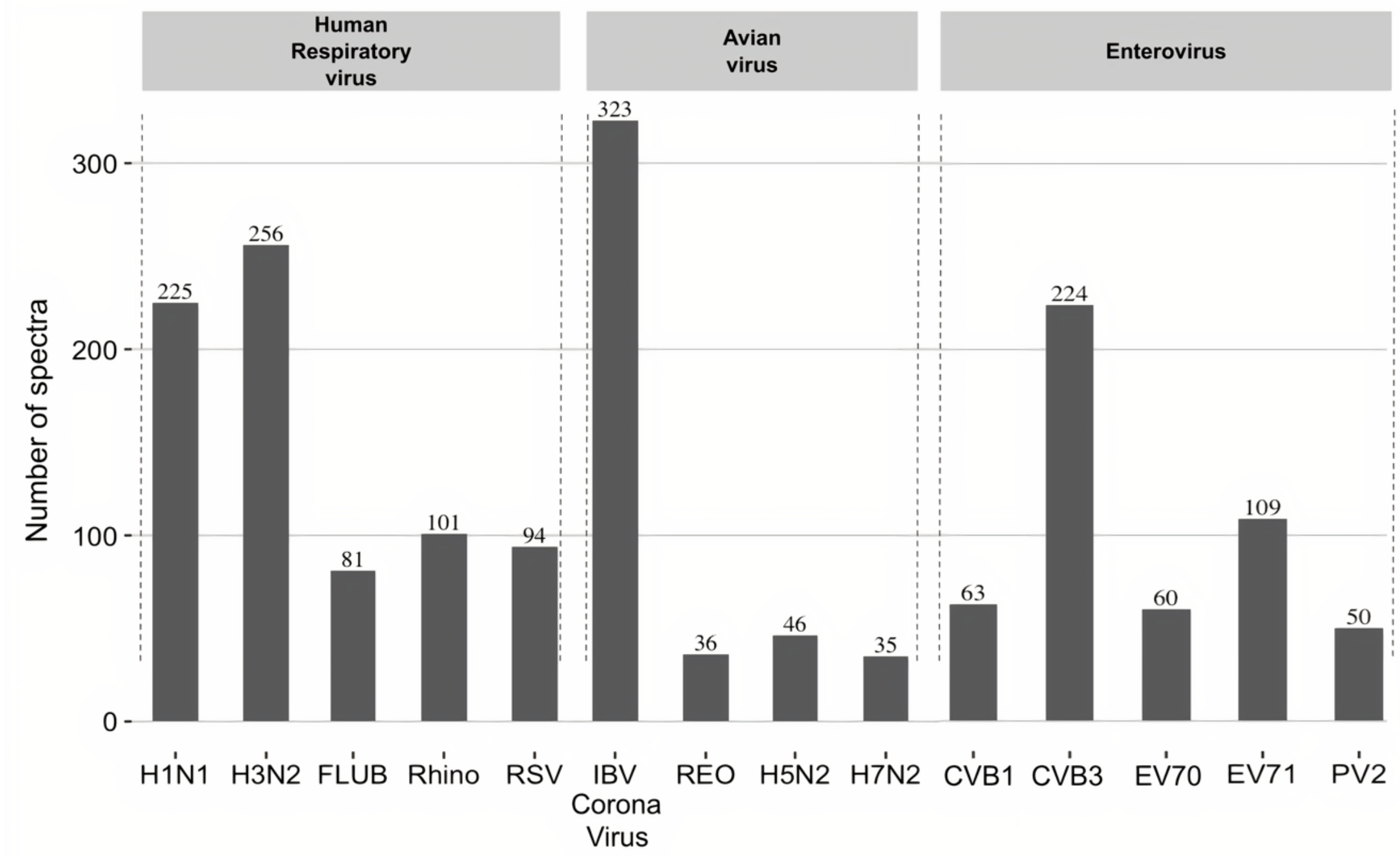

**Figure 3.** Number of spectra in our dataset for Human respiratory viruses, avian viruses, and Human enteroviruses. H1N1, H3N2, H5N2, and H7N2 are subtypes of the influenza A virus; FLUB: influenza B virus; Rhino: rhinovirus; RSV: respiratory syncytial virus; IBV: infectious bronchitis virus; Reo: reovirus; CVB1 and CVB3: Coxsackievirus B type 1 and 3; EV70 and EV71: enteroviruses. Numbers above each column indicate the number of spectra collected for each virus. These spectra all have ground truth labels which are the virus type/subtype. Note that, for classification tasks, we apply data augmentation to add more samples to virus classes that have fewer number

of spectra samples so that for each classification task, every virus type has an equal number of spectra samples in the training set.

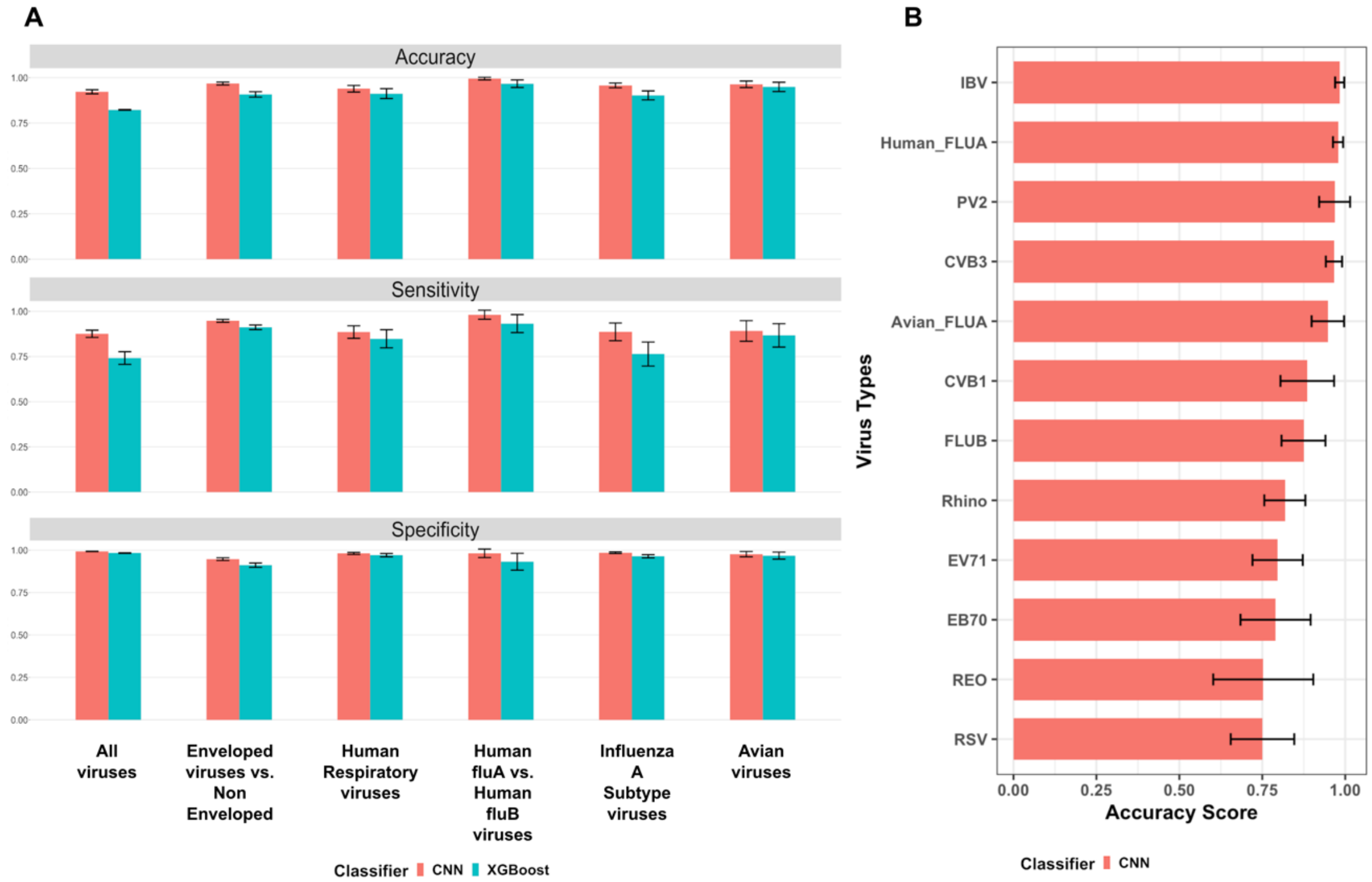

**Figure 4. A.** The classification performance of our CNN model and the XGBoost model on 6 experiments: 1) All viruses (classification of all virus types): Avian, Enteroviruses, Human Respiratory viruses; 2) Enveloped viruses vs. Non-Enveloped: FLUA and FLUB, IBV Coronavirus, RSV are Enveloped, Reovirus, Enterovirus CVB1/CVB3/EV70/EV71/PV2, Rhino are Non-Enveloped; 3) Human Respiratory viruses; 4) Human FLUA vs. Human FLUB viruses; 5) Influenza A subtypes; and 6) Avian viruses. Three metrics (accuracy, sensitivity, and specificity) are measured for both classification models. Results for all metrics are obtained by running a 5-fold cross-validation 5 times for fair comparison (each error bar represents the standard deviation of the corresponding metric score for each experiment across 5-fold cross-validation in 5 tests); **B.** Accuracy score for every virus type in the All-viruses classification task (each error bar represents the standard deviation of the corresponding accuracy score for each virus type across 5-fold cross-validation in 5 tests).

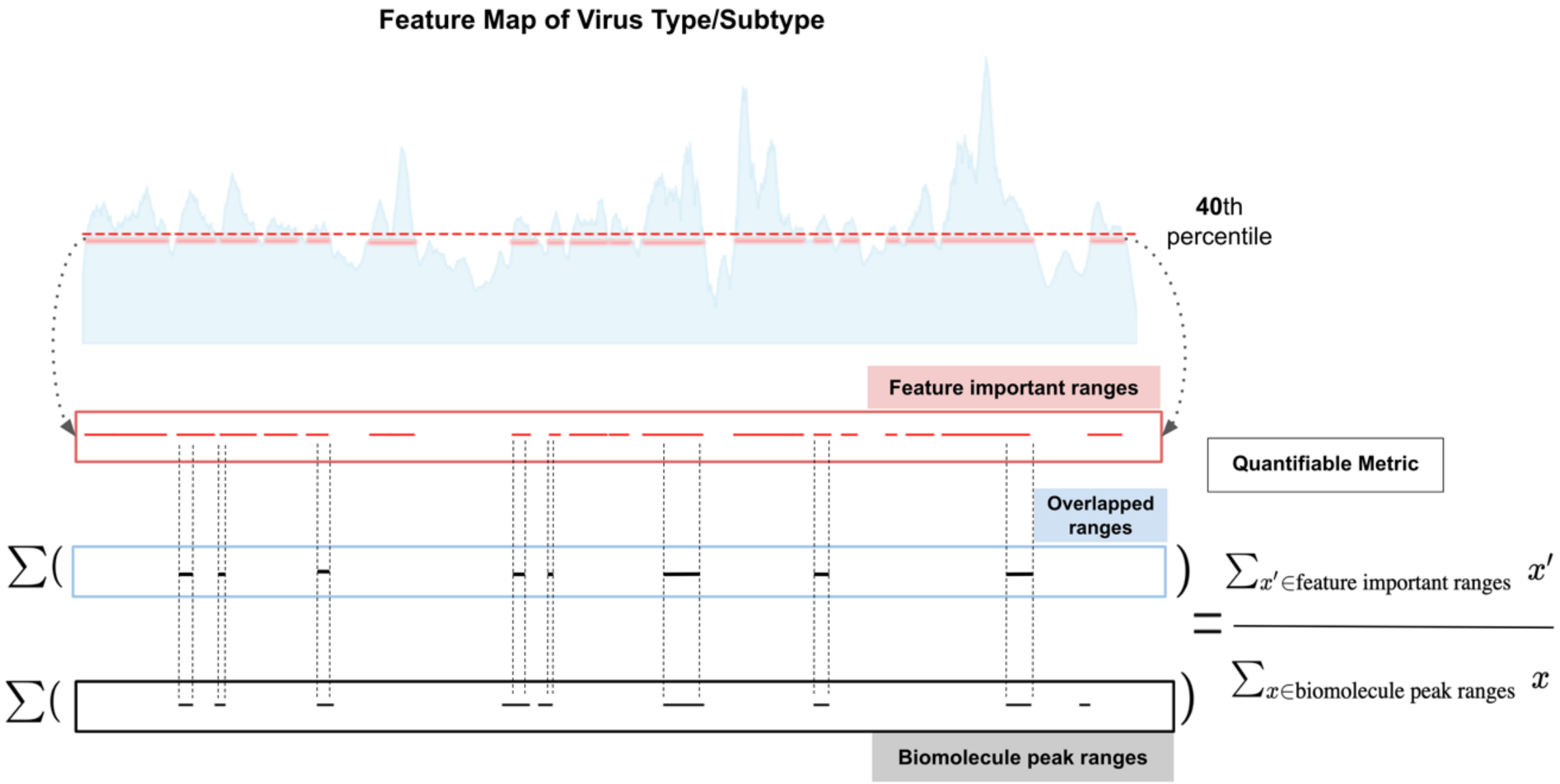

**Figure 5.** Illustration of the quantifiable matching score calculation leveraging biomolecule peak ranges and important ranges extracted from ML-calculated feature maps of each virus type (or subtype, depending on the classification task). A threshold of 40th percentile is applied to the ML-calculated feature importance map so that Raman bands with importance scores below the threshold are discarded, and the remaining wavenumbers above the threshold are considered as important Raman ranges for identifying the virus based on ML and can then be correlated with biomolecule peak ranges.

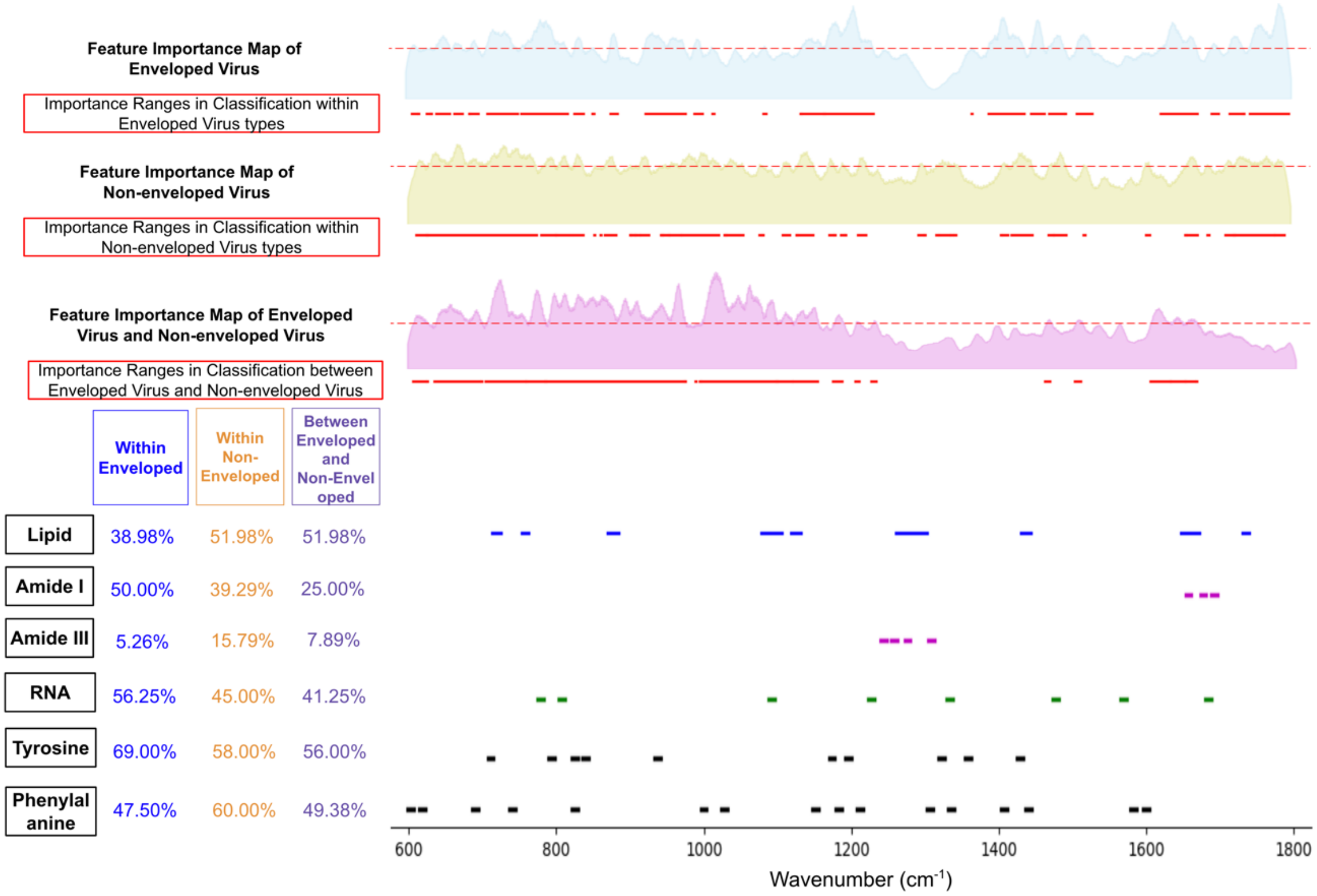

**Figure 6.** Biomolecule peak ranges, ML-calculated feature importance map, and important Raman ranges (above 40th percentile threshold) for classification experiments: 1) Within enveloped virus types (Avian FLUA, IBV Coronavirus, Human FLUA, Human FLUB, RSV); 2) Within non-enveloped virus types (Enterovirus (CVB1, CVB3, EV70, EV71, PV2), Rhino, Reovirus); 3) Between Enveloped and Non-enveloped viruses. Feature importance maps are extracted from intermediate layers of the CNN as described in Fig. 1*B*. The matching score for each classification experiment is calculated by correlating ML-selected important ranges with each biomolecule's known Raman peak ranges. (See *SI Appendix,* Fig. S6 for matching scores with more functional groups)

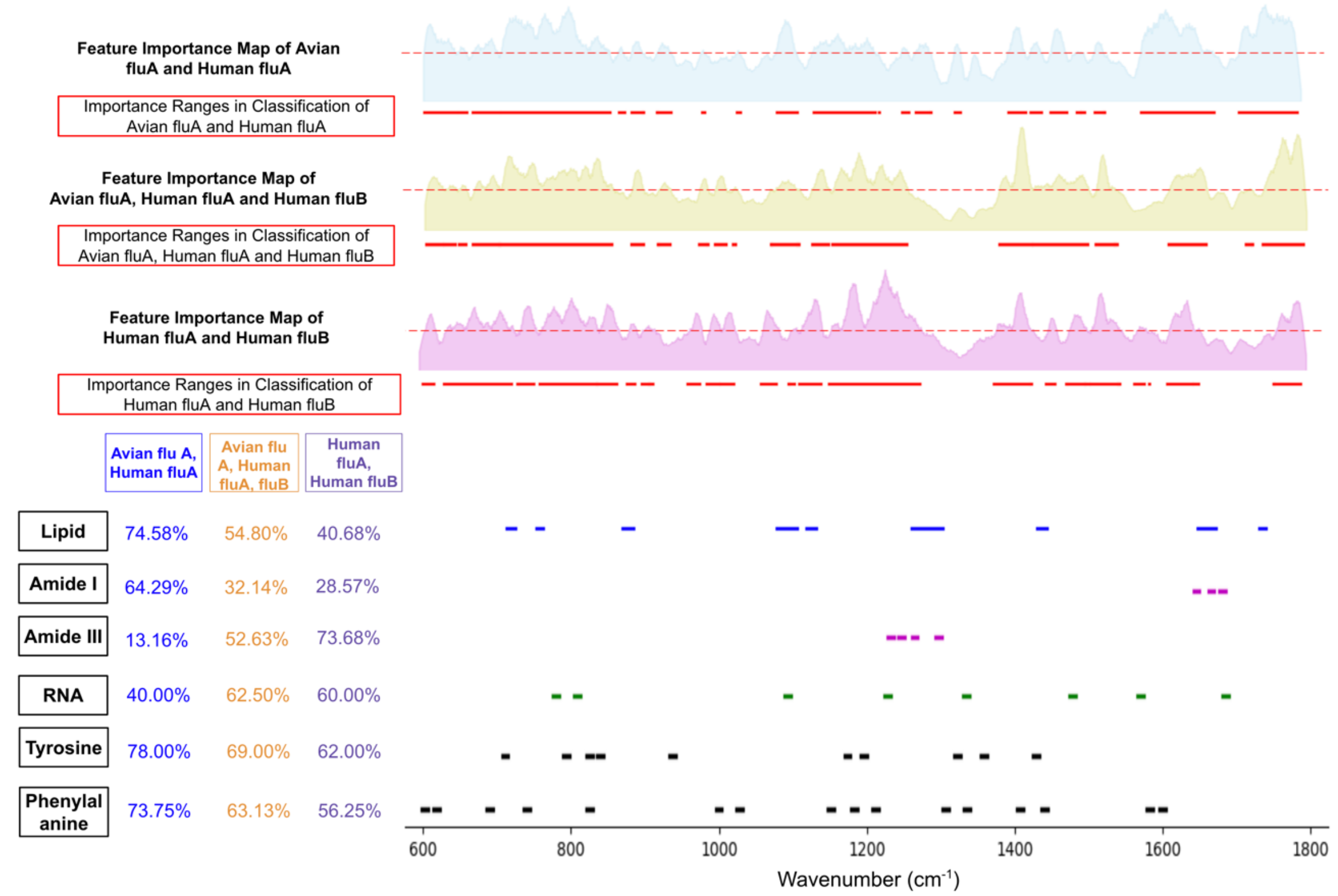

**Figure 7.** ML-calculated feature importance map, and important Raman ranges for classification experiments: 1) Avian FLUA vs. Human FLUA; 2) Avian FLUA, Human FLUA and Human FLUB; 3) Human FLUA and Human FLUB. (See *SI Appendix*, Fig. S5 for matching scores with more functional groups).

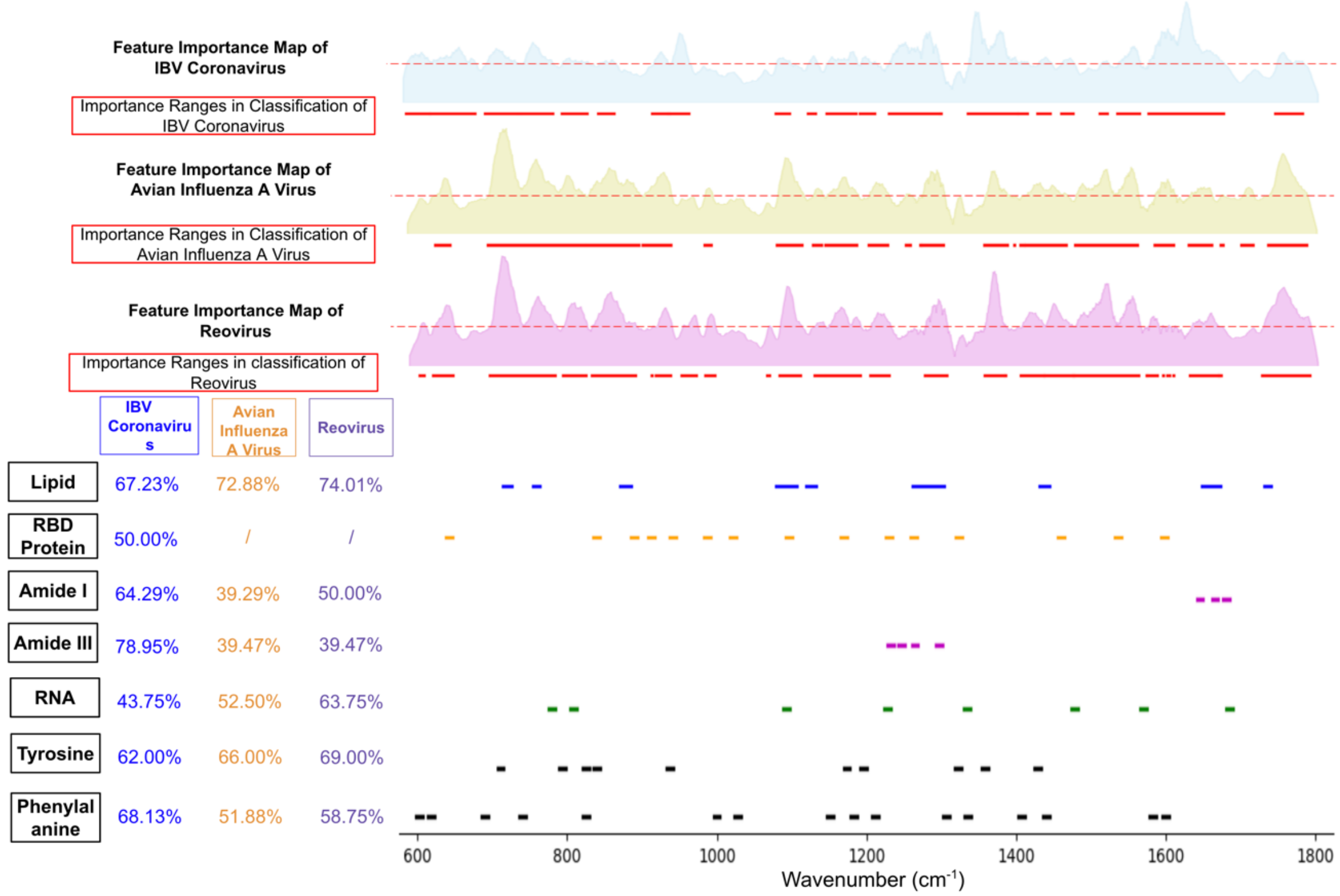

**Figure 8.** ML-calculated feature importance map, and important Raman ranges for classifying three types of avian viruses: IBV Coronavirus, Avian Influenza A virus and Reovirus. Feature important maps and matching scores are given for each avian virus type. The matching score for RBD Protein only applies when correlating with IBV Coronavirus because RBD Protein is an exclusive biomolecule in IBV. (See *SI Appendix*, Fig. S2 for matching scores with more functional groups).